\documentclass[10pt, preprint]{emulateapj} 		
\usepackage{color}

\usepackage{style_wei_symbols}
\usepackage{style_wei}
\usepackage{style_wei_journal_names}
  \renewcommand{\lrsp}{{Liv.~Rev.~Solar Phys.}}

\usepackage{natbib}
\def\referee#1{{#1}}
\def\revise#1{{#1}}

\begin{document}

\title{First High-resolution Spectroscopic Observations of an Erupting Prominence Within a Coronal Mass Ejection
by the {\it Interface Region Imaging Spectrograph} ({\it IRIS})}

\author{Wei Liu\altaffilmark{1}$^,\,$\altaffilmark{2},		
Bart De Pontieu\altaffilmark{1}$^,\,$\altaffilmark{3}, 
Jean-Claude Vial\altaffilmark{4}, Alan M.~Title\altaffilmark{1}, 
Mats Carlsson\altaffilmark{3}, Han Uitenbroek\altaffilmark{5},
Takenori J.~Okamoto\altaffilmark{6}, Thomas E.~Berger\altaffilmark{7}, 
Patrick Antolin\altaffilmark{8}}

\altaffiltext{1}{Lockheed Martin Solar and Astrophysics Laboratory, 
  3251 Hanover Street, Palo Alto, CA 94304, USA; weiliu@lmsal.com}	
\altaffiltext{2}{W.~W.~Hansen Experimental Physics Laboratory, Stanford University, Stanford, CA 94305, USA}
\altaffiltext{3}{Institute of Theoretical Astrophysics, University of Oslo, P.O. Box 1029, Blindern, N-0315, Oslo, Norway}
\altaffiltext{4}{Institut d'Astrophysique Spatiale, Universit\'{e} Paris XI/CNRS, 91405, Orsay Cedex, France}
\altaffiltext{5}{National Solar Observatory, P.O.~Box 62, Sunspot, NM 88349, USA}
\altaffiltext{6}{Institute of Space and Astronautical Science, Japan Aerospace Exploration Agency, Yoshinodai, Sagamihara, 252-5210, Japan}
\altaffiltext{7}{NOAA Space Weather Prediction Center, Boulder, CO, USA}
\altaffiltext{8}{National Astronomical Observatory of Japan, 2-21-1 Osawa, Mitaka, Tokyo 181-8588, Japan}

\shorttitle{IRIS Observations of a Fast Prominence Eruption}
\shortauthors{Liu et al.}
\slugcomment{Accepted by ApJ, February 15, 2015}

\begin{abstract}	

Spectroscopic observations of prominence eruptions associated with coronal mass ejections (CMEs), 
although relatively rare, can provide valuable plasma and 3D geometry diagnostics.
We report the first observations by the {\it Interface Region Imaging Spectrograph} ({\it IRIS}) mission
of a spectacular fast CME/prominence eruption associated with an equivalent X1.6 flare on 2014 May 9.
The maximum plane-of-sky and Doppler velocities of the eruption are 1200 and $460 \kmps$, respectively.
There are two eruption components separated by $\sim$$200 \kmps$ in Doppler velocity: 
a primary, bright component and a secondary, faint component, 
suggesting a hollow, rather than solid, cone-shaped distribution of material.
The eruption involves a left-handed helical structure undergoing counter-clockwise
(viewed top-down) unwinding motion.
There is a temporal evolution from upward eruption to 
downward fallback with less-than-free-fall speeds and decreasing nonthermal line widths.
We find a wide range of \mgiik/h line intensity ratios 
(less than $\sim$2 expected for optically-thin thermal emission): 
the lowest ever-reported median value of 1.17 found in the fallback material 
and a comparably high value of 1.63 in 	
nearby coronal rain and intermediate values of 1.53 and 1.41 in the two eruption components.
The fallback material exhibits a strong ($>$$5 \sigma$) linear correlation between the k/h ratio and the Doppler velocity
as well as the line intensity.
We demonstrate that Doppler dimming of scattered chromospheric emission 
by the erupted material can potentially explain such characteristics.



\end{abstract}


\section{Introduction}
\label{sect_intro}



As one of the major drivers of space-weather disturbances, coronal mass ejections 
\citep[CMEs; see][for recent reviews]{ChenPF.CME.review.2011LRSP....8....1C, WebbD.HowardT.CME.obs.review.2012LRSP....9....3W}
are powerful eruptions of typically $10^{15}$\,--\,$10^{16} \g$ masses from the Sun
traveling at 100\,--\,$2000 \kmps$ and involving energies of $10^{30}$\,--\,$10^{32} \erg$.
CMEs seen in white light often exhibit a classic three-part morphology 
\citep[e.g.,][]{HundhausenA.SMM.CME.def.1984JGR....89.2639H, SchwennR.CME.terminology.1996Ap&SS.243..187S}
consisting of a bright leading front surrounding a dark cavity containing a bright core. 
The core is identified as an erupting prominence \citep[e.g.,][]	
{Tandberg-HanssenE.prominenceBook.1995nsp..book.....T, MartinSF.prominence-review.1998SoPh..182..107M,
Labrosse.prominence-review.2010SSRv..151..243L, MackayDH.prominence-review.2010SSRv..151..333M,
ParentiS.prom.review.2014LRSP...11....1P}, 
some cool ($T$$\sim$$10^4 \K$) and dense ``chromospheric" plasma 		
residing in the hot ($T$$\sim$$10^6 \K$) and tenuous corona prior to its eruption.
	%

Prominence eruptions and CMEs in general are conventionally detected with {\it global imaging instruments},
which can constrain the morphology and plane-of-sky (POS) velocity among other observables.
Such observations are obtained either monochromatically
(e.g., in \Ha, extreme ultraviolet [EUV], or white light; 
\citealt{StCyrC.LASCO.CME.stat.2000JGR...10518169S}) 
or in multiple passbands 
(e.g., with the Coronal Multichannel Polarimeter [CoMP];
\citealt{TianH.CoMP.CME.2013SoPh..288..637T}).

{\it Spectroscopic observations} of prominence eruptions or CMEs within restricted fields of view (FOVs) of slit spectrographs, 
although comparably rare, can provide valuable, complementing plasma diagnostics, 
e.g., density, temperature, and Doppler velocity.
Such observations were obtained (i) in the outer corona (1.4\,--\,$10 \Rsun$)
by the UltraViolet Coronagraph Spectrometer \citep[UVCS; e.g.,][]
{CiaravellaA.1st.UVCS.CME.1997ApJ...491L..59C, RaymondJ.UVCS.fast.CME.2003ApJ...597.1106R} 	
onboard the {\it Solar and Heliospheric Observatory} (\soho),
and (ii) in the inner corona ($r \lesssim 1.5 \Rsun$) by spectroheliographs on {\it Skylab} 
\citep[e.g.,][]{SchmahlE.90perc.prom.mass.drain.in.CME.1977SoPh...55..473S, WidingK.Skylab.erupt.prom.1986ApJ...308..982W},	
the Ultraviolet Spectrometer and Polarimeter (UVSP) on the {\it Solar Maximum Mission} 
\citep[{\it SMM}; e.g.,][]{FontenlaJ.SMM.UVSP.prom.eruption.1989SoPh..123..143F},
the Solar Ultraviolet Measurements of Emitted Radiation (SUMER) and
Coronal Diagnostic Spectrometer (CDS) on \soho\ \citep[e.g.,][]{WiikJ.SUMER.CDS.erupt.prom.CME.1997SoPh..175..411W},
the EUV Imaging Spectrometer (EIS) on \hinode\ (e.g., \citealt{HarraL.CME.dimming.Hinode.EIS.2007PASJ...59S.801H, 
JinM.EIS.CME.outflow.2009ApJ...702...27J, TianH.spectr.obs.CME.2012ApJ...748..106T}; Williams et al. 2015, in prep.),
and ground-based instruments \citep[e.g.,][]{PennM.erupt.prom.CDS.KPVT.2000SoPh..197..313P}.
Notably, during 1996\,--\,2005, UVCS detected $>$1000 CMEs \citep{GiordanoS.UVCS.CME.catalog.2013JGRA..118..967G}, 
about 10\% of the CMEs detected by the \soho\ Large Angle and Spectrometric Coronagraph (LASCO) during the same period.
The most commonly detected feature of UVCS CMEs (in $\sim$70\% of them) is the cool prominence material.



The recently launched {\it Interface Region Imaging Spectrograph} mission 
\citep[\iris;][]{DePontieuB.IRIS.mission.2014SoPh..289.2733D},
thanks to its advanced capabilities over previous generations of instruments, 
has provided a new opportunity to observe CMEs and especially prominence eruptions in unprecedented detail.
\iris\ offers a high resolution of	
$0\farcs 33$\,--\,$0\farcs4$ in space, $2 \s$ in time, and $1 \kmps$ in Doppler velocity.
A unique advantage of \iris\ is the combination of spectra
and slit-jaw images (SJIs), which provides simultaneous Doppler and POS velocity measurement
and can thus constrain the true three-dimensional (3D) geometry and velocity vector.
Its ultraviolet (UV) spectra cover a wide temperature range of 	
$5\E{3}$\,--\,$10^7 \K$ and are particularly sensitive to chromospheric and transition-region temperatures
of prominence material. 
With its $175 \arcsec \times 175 \arcsec$ FOV covering the low corona,
\iris\ can provide critical clues to the initiation and early development of a CME/prominence eruption
and complement instruments covering the high corona 
(e.g., coronagraphs and UVCS which was turned off on 2012 December 15).

We report in this paper the first result of \iris\ observations of an erupting prominence
within a CME. We present the overview of observations in \sect{sect_obs}, 
data analysis results in \sect{sect_analysis}, and concluding remarks in \sect{sect_conclude}.
%
%
 \begin{figure*}[thbp]      
 \includegraphics[height=7in]{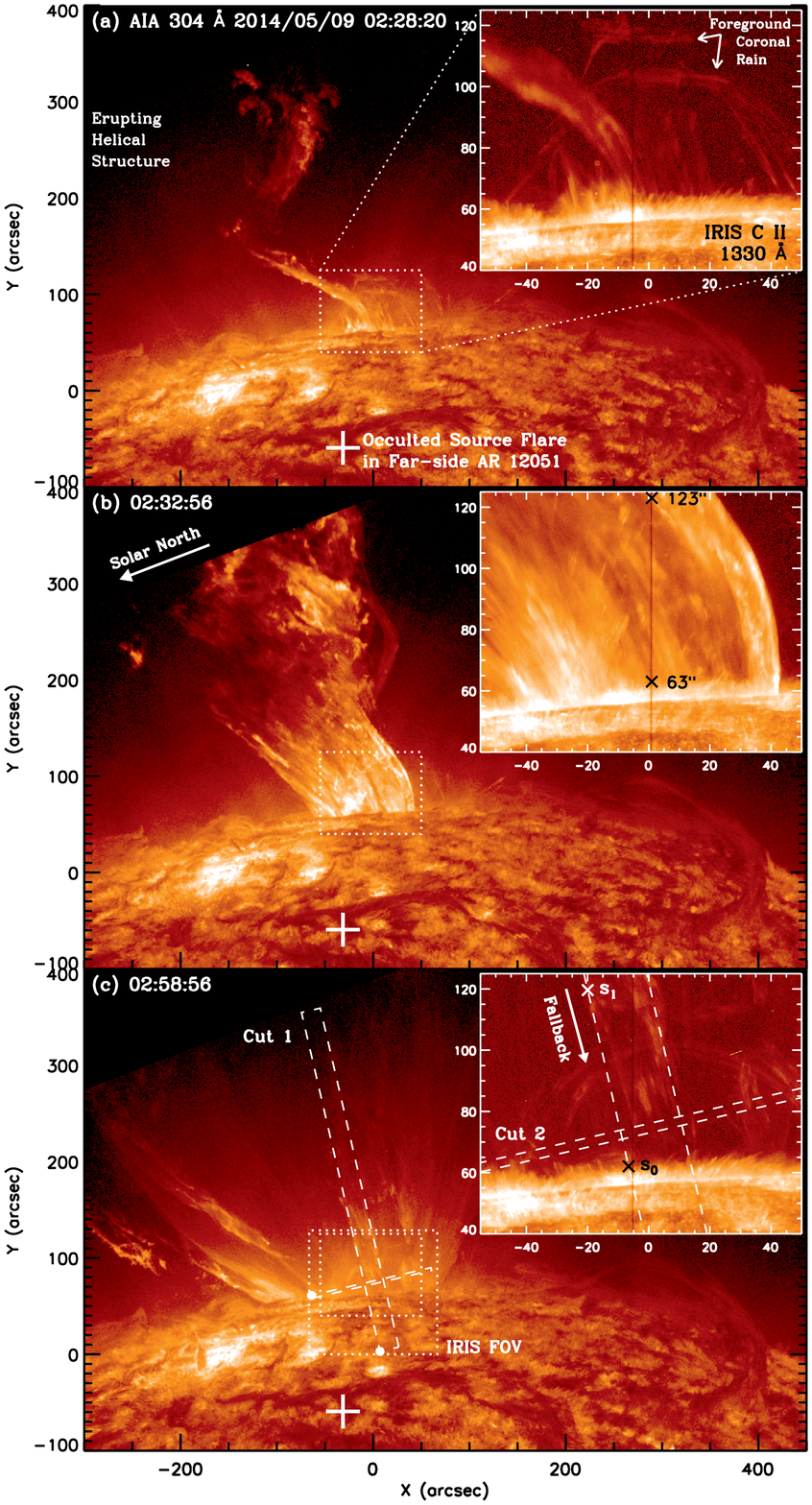}	
 \includegraphics[height=7in]{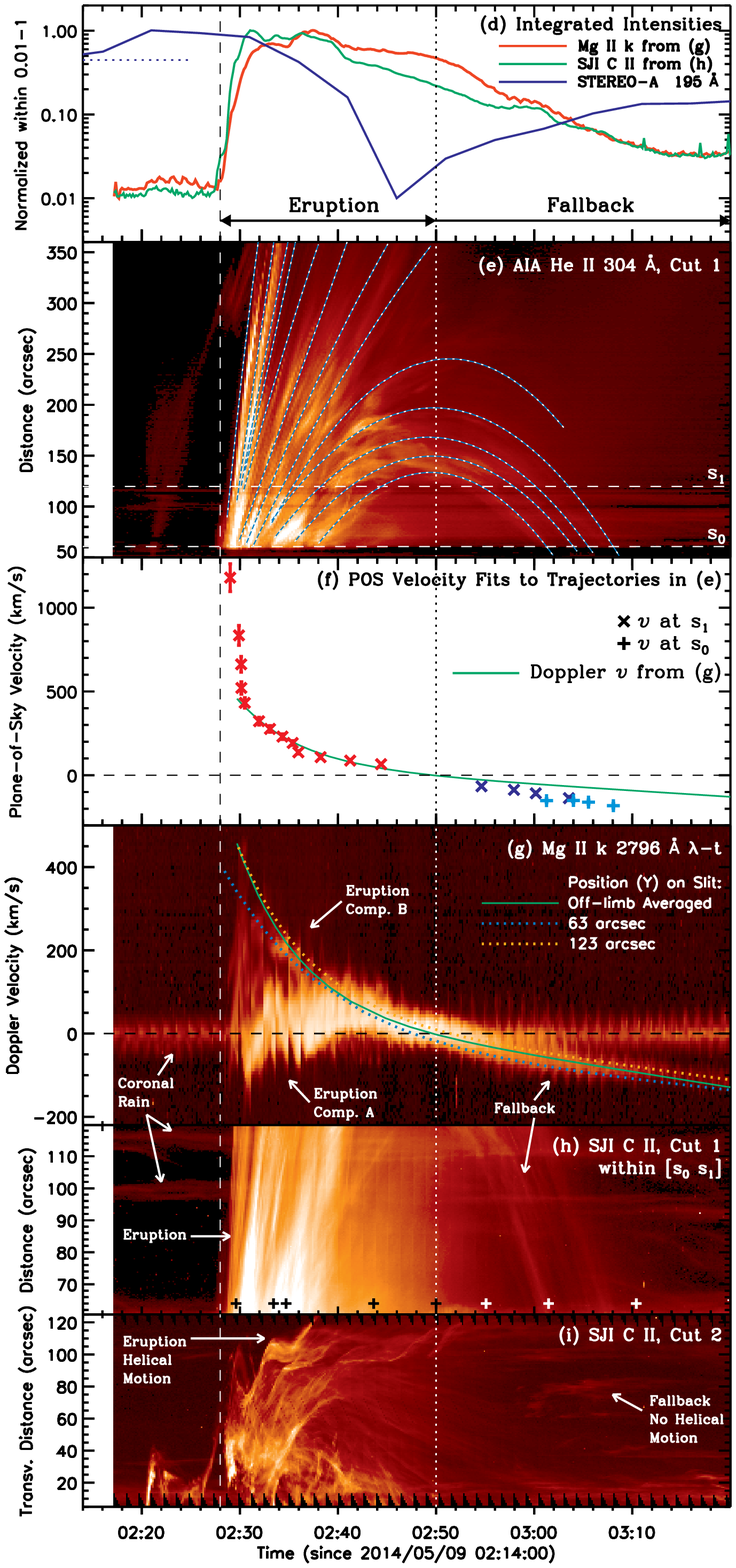}	
 \caption[]{\small
 Images showing milestones of the CME/prominence eruption (left) and history of various quantities (right).
 (a)--(c) \sdo/AIA 304~\AA\ images rotated and shifted to match the \iris\ orientation,
 with the origin of (X, Y) located at the center of the bottom edge of the \iris\ FOV.
 (The accompanying online Animation~1 is oriented with Solar West up.)
 The insets show simultaneous \iris\ \cii\ 1330~\AA\ SJIs	
 of the region within the small white box. The large box in (c) indicates the original \iris\ FOV.
 The two narrow rectangles are cuts used to obtain space--time plots in (e), (h) and (i),
 with the white filled circle denoting the start of the cut.
 The large plus sign shows the location of the source flare centroid detected by \stereo-A behind the limb.
  (d) Various light curves: 
 \iris\ \mgiik\ 2796~\AA\ (red) and \cii\ 1330~\AA\ SJI (green) intensities obtained by vertically collapsing 
 the corresponding space--time plots in (g) and (h), respectively, 	
 and \stereo-A EUVI 195~\AA\ full-Sun flux (blue) with the pre-event background (blue-dotted line) subtracted.
  (e) AIA 304~\AA\ space--time plot obtained from Cut~1 (along the central axis of the ejection) shown in (c),
 overlaid with parabolic fits to trajectories shown in dotted lines.
 $s_0$ and $s_1$ here and in (c) mark the off-limb spatial range of the cut within the \iris\ FOV. 
 The velocities of the fitted trajectories at these two positions are shown in (f).
  (g) \mgiik\ $\lambda$--time plot by averaging spectra over all off-limb positions of the slit.
 The green curve here (and reproduced in (f)) is a parametric fit to the temporal transition
 from red- to blueshifts.	
 Its counterparts in dotted lines are obtained from two off-limb positions (as marked in (b)),		
 showing similar temporal evolution and indicating
 larger redshifts or smaller blueshifts at greater heights at any given time (as also shown in \fig{map-mg.eps}(h)).
  (h) and (i) \iris\ \cii\ 1330~\AA\ SJI space--time plots obtained from Cuts~1		
 and 2.		
 The latter shows sinusoidal transverse displacements, manifesting unwinding motions of helical threads.
 The plus signs in (h) mark the times for the \mgiik/h ratio analysis shown in 
 the bottom panels of \fig{ratio.eps}.
 } \label{overview.eps}
 \end{figure*}
%

\section{Overview of Observations}
\label{sect_obs}


	%

On 2014 May 9, a major eruption involving a flare and a CME with a prominence eruption	
occurred in NOAA active region (AR) 12051 on the far side of the Sun. 
The flare kernel, observed by the Extreme UltraViolet Imager (EUVI) on 
the {\it Solar TErrestrial RElations Observatory} Ahead spacecraft (\stereo-A), 	
was located $31 \degree$ behind the west limb seen from the Earth perspective.
Its POS projection, shown as the plus sign in \fig{overview.eps}(a), 
was located at ($815 \arcsec$, $-204 \arcsec$), $110 \arcsec$ inside the limb.	
The pre-event background removed full-disk EUVI 195~\AA\ flux 
(\fig{overview.eps}(d), blue line, at a 5~min cadence)
peaked at 02:21~UT and $1.16\E{7} {\rm photons} \, \ps$, which translates to an equivalent
\goes\ X-ray flare class of X1.6, according to empirical scaling 
\citep{NittaN.STEREO.behind-limb-flare-class.2013SoPh..288..241N}.
The CME observed by \soho/LASCO had a maximum POS speed of $1300 \kmps$, according to the
CACTUS catalog (http://sidc.oma.be/cactus).

As seen by the Atmospheric Imaging Assembly (AIA) on the {\it Solar Dynamics Observatory} (\sdo) 
in the \heii~304~\AA\ channel, the prominence eruption occurred in two episodes starting at 02:20 and 02:28~UT,
with leading-edge speeds of 1060 and $1180 \kmps$, respectively. 
Episode~1 involved a single helical material thread 
expanding radially (\fig{overview.eps}(a)).
Episode~2 was more complex, consisting of sub-structured material bundles 
spanning a cone shape up to $\sim$$90 \degree$ in angular extent (\figs{overview.eps}(b) and (c)), 
whose central axis (along ``Cut~1") was oriented $10 \degree$ counter-clockwise from the radial 
direction.		
Episode~2 lasted more than an hour up to 03:45~UT, with material being 
episodically ejected in some directions and at the same time falling back to the Sun in 
some other directions. 

\iris\ was running 	
large coarse 8-step rasters with a step size of $2\arcsec$ and 9.6~s cadence (8~s exposure)
through 03:21~UT. 	
SJIs in a single \cii~1330~\AA\ channel (bandwidth: 55 \AA) covered a
maximum $133 \arcsec \times 129 \arcsec$ FOV, as shown in \fig{overview.eps}(c).
The \iris\ slit, oriented at $13 \degree$ clockwise from the eruption axis,
missed the first episode but fully covered the second,
which will be the focus of this paper.
\tab{table_timeline} summarizes the milestones of this event and corresponding coverage
by different instruments.
%
\begin{table}[bthp]	
\scriptsize		
\revise{\caption{Event Milestones on 2014 May 09.}}  
\tabcolsep 0.03in	
\begin{tabular}{ll}
\tableline \tableline
 02:16/02:21   &  Flare onset/peak seen by \stereo-A (5 min cadence) \\
 02:18         &  EUV wave onset seen by \sdo/AIA \\
 02:20--02:28  &  Prominence eruption Episode~1 seen by \sdo/AIA \\
               &  Max.~POS speed $1060 \kmps$ at $\sim$02:21 \\
 02:28--03:45  &  Prominence eruption Episode~2 seen by \sdo/AIA \\
               &  Max.~POS speed $1180 \kmps$ at $\sim$02:29 \\
 02:28--03:21  &  \iris\ slit coverage of Episode~2 eruption \\
               &  Max.~redshift $460 \kmps$ at 02:31  \\
 02:48         &  CME seen by \soho/LASCO (24 min cadence) \\
 02:50         &  Transition from eruption to fallback within \iris\ slit \\

\tableline  \end{tabular}

\label{table_timeline} \end{table}
%

\section{\iris\ Data Analysis}
\label{sect_analysis}

\referee{\subsection{Wavelength Calibration}
\label{subsect_calib}
}

\referee{We performed on \iris\ level-2 data {\it relative} (not {\it absolute}) wavelength calibration,
which practically served our purpose of measuring the Doppler velocity of the eruption.
Shown as the vertical dotted line in \fig{map-3w.eps}, the reference wavelength 
for the Doppler velocity in each spectral line window
was selected at the centroid of the corresponding line profile averaged over 
the on-disk portion of the slit through this event (02:17--03:21~UT).
Therefore, all the Doppler velocities are measured with respect to the quiet-Sun region near the limb
within the \iris\ slit, which is located in the neighborhood of the eruption source region 
(the plus sign in \fig{overview.eps}(a)).} 	

\referee{The absolute uncertainty of this relative wavelength calibration can be estimated as follows.
Taking the \mgiik~2796~\AA\ line as an example, 
the sources of uncertainty include
(i) the $20 \mA$ wavelength shift (equivalent to $2 \kmps$) of this line measured near
the solar limb from that at the disk center \citep{KohlJL.Parkinson.MgII.center-2-limb.1976ApJ...205..599K}
and (ii) the \iris\ orbital thermal variation of $\sim$$3 \kmps$
\citep{DePontieuB.IRIS.mission.2014SoPh..289.2733D}.
These errors combined give an overall uncertainty of $\sim$$4 \kmps$
for the absolute Doppler velocity, two orders of magnitude smaller
than the typical values of hundreds of $\kmps$ measured in this eruption.
}

\subsection{Temperature Distribution}	
\label{subsect_tem}



The eruption was observed in \cii~1330~\AA\ SJIs and spectra of cool lines at chromospheric to transition region
temperatures, including \mgii\ ($T$$\sim$$10^{4} \K$), \cii\ ($10^{4.3} \K$), 
\siiv\ ($10^{4.8} \K$), and \oiv\ ($10^{5.2} \K$), as shown in \fig{map-3w.eps},
corresponding to the erupting prominence material.	
\iris\ detected no eruption-associated signal in the
\fexii~1349~\AA\ ($10^{6.2} \K$) or \fexxi~1354~\AA\ ($10^{7.0} \K$) line, although AIA detected
in its \fexii~193~\AA\ and \fexiv~211~\AA\ ($\sim$$10^{6.3} \K$) channels
a dome-shaped, CME-generated EUV wave \citep[e.g.,][]{LiuW.OfmanL.EUV.wave.review.2014SoPh..289.3233L}
appearing above the limb at 02:18~UT preceding the prominence eruption.	
We found no obvious delay among line intensities at different temperatures at a given spatial location
and Doppler velocity, indicating no detectable temperature change of the erupting material
passing through the \iris\ slit.

\subsection{Velocity Distribution: Plane-of-Sky and Doppler}
\label{subsect_vel}



We selected a $20 \arcsec$ wide Cut~1 along the eruption axis to
cover a large portion of the $14 \arcsec \times 129 \arcsec$ region scanned
by the \iris\ slit and to obtain an AIA 304~\AA\ space--time plot shown in \fig{overview.eps}(e).
The mass ejections exhibit ballistic-shaped trajectories, which were fitted 
with parabolic functions shown as dotted lines. 
We found a cascade of the ejections toward lower velocities at later times.
As shown in \fig{overview.eps}(f), the velocity at the top of the \iris\ FOV (labeled $s_1$ in panel~(e)), 
starts with $1180 \pm 120 \kmps$ at 02:29:01~UT and rapidly decreases to $430 \pm 40 \kmps$ at 02:30:30~UT,
followed by a gradual drop to $66 \pm 6 \kmps$ at 02:44:24~UT.
Ejections with speeds $\lesssim$$200 \kmps$ at $s_1$ turn back and start falling around 02:50~UT.
In general, faster ejections reach greater heights and return to the \iris\ FOV later at higher speeds.
\fig{overview.eps}(h) shows an \iris\ \cii~1330~\AA\ SJI space--time plot from the same Cut~1 at higher resolution
with similar behaviors because of \referee{its similar sensitivity (as \heii~304~\AA) 
to the ejected prominence material at typically transition-region/chromospheric temperatures
\citep[see][for more information on the \cii~1330~\AA\ formation temperature]
{AvrettE.IRIS-lines.Mg.C.Si.2013ApJ...779..155A}.}

\begin{figure*}[thbp]      
 \epsscale{1.1}
 \plotone{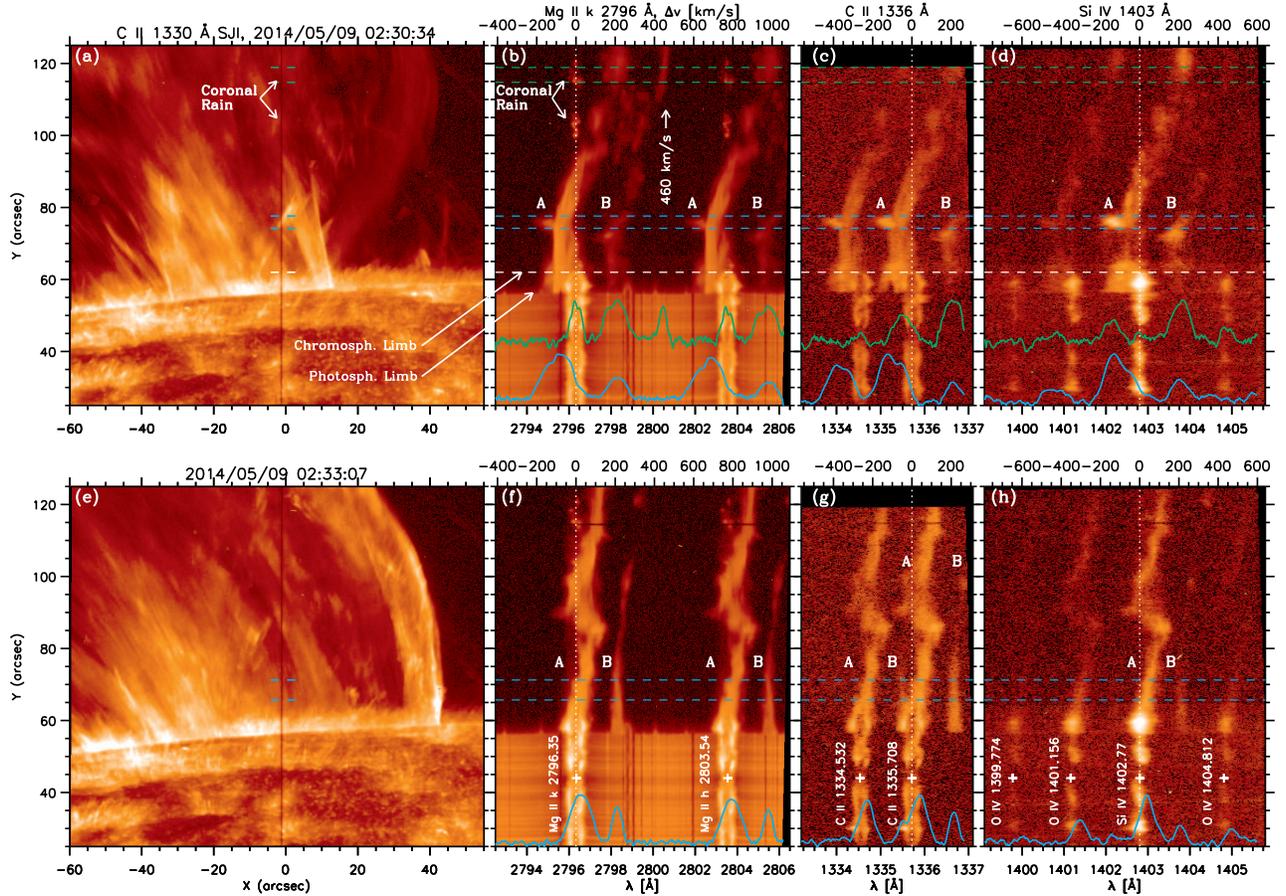}	
 \caption[]{	
 \cii~1330~\AA\ SJI (left) and simultaneous spectra in three line windows (right; see Animation~2), 
 \mgiik\ 2796~\AA, \cii\ 1336~\AA, and \siiv\ 1403~\AA, for two selected times (top and bottom).
 The highest Doppler shift in the eruption of $460 \kmps$ is identified in (b).
 In each spectral panel, the top $x$-axis shows the Doppler velocity, whose reference wavelength
 \referee{(see text in \sect{subsect_calib})} is marked by the vertical dotted line. 
 The plus signs in the bottom panels denote the nominal rest wavelengths of bright lines. 
 Each pair of horizontal dashed lines define the height range for the spectrum to be
 averaged to obtain the logarithmic profile shown at the bottom in the corresponding color.
 The horizontal white dashed line in the top panels indicates the height of the chromospheric limb (transition region)
 at the slit position shown on the left. ``A" and ``B" label the two components (primary and secondary)
 of the eruption in velocity space.
 } \label{map-3w.eps}
 \end{figure*}

In the Doppler velocity distribution observed by \iris, a striking feature is a {\it two-component composition}
during the early phase of the eruption:
a primary (A) bright and broad component being slightly blueshifted and/or mildly redshifted,
and a secondary (B) highly redshifted, comparably faint and narrow component, as shown in \fig{map-3w.eps}.
The two components, yet with different structures, are somewhat parallel to each other, 
separated by $\sim$$200 \kmps$, with a similar trend of growing redshifts (up to $460 \kmps$) with height.
They can be identified in all bright lines, including the \mgiik~2796~\AA\ and h~2803~\AA,
\cii~1335 and 1336~\AA, and \siiv~1403~\AA\ lines. (Note that the \cii~1335~\AA\ component~B
almost overlaps with the 1336~\AA\ component~A.) 
One possibility consistent with the gap between the two components is that the erupting material
is spatially distributed in a {\it hollow}, rather than {\it solid}, cone shape.


As time progresses, the two components gradually merge into a single component with decreasing line widths.
At later times, the entire spectrum evolves from red- to blueshifts, while maintaining the same
general slope with height as noted below (see \fig{map-mg.eps}, bottom).
This evolution can be clearly seen from the wavelength-- or velocity--time plot shown in \fig{overview.eps}(g),
obtained by averaging the \mgiik~2796~\AA\ spectra over the entire region 	
above the chromospheric limb. 
We fitted the velocity--time positions of component B and later the merged single component
with a smooth spline function, shown as the solid green curve, to characterize the temporal evolution.
This curve exhibits an initially rapid and then gradual decline with time,
similar to that of the POS velocity shown in \fig{overview.eps}(f).
More interestingly, its crossing at zero Doppler velocity, i.e., the change from red- to blueshifts
at 02:50~UT (marked by the vertical dotted line),
{\it coincides with} the switch of sign for the POS velocity and the apexes of the ballistic POS trajectories
shown in \fig{overview.eps}(e), indicating a transition from upward eruption to downward fallback of the material.
This transition may also explain the onset of the faster drop of the \mgiik~2796~\AA\ intensity at the same time
(see \fig{overview.eps}(d), red line).
\referee{The uncertainty of this temporal coincidence is estimated at $\pm 2$~minutes,
bracketed by the zero-crossings of the two dotted lines in \fig{overview.eps}(g), which are
the counterparts of the solid curve obtained near the bottom and top edges
of the off-limb portion of the \iris\ slit 
(at $Y= 63\arcsec$ and $123 \arcsec$, see \fig{overview.eps}(b)).
This conservative error estimate can also account for the temporal spread of the turning points
of the different POS trajectories shown in \fig{overview.eps}(e) and the $\sim$$4 \kmps$
uncertainty in the absolute Doppler velocity noted in \sect{subsect_calib}.}

\referee{We also note in \fig{overview.eps}(g) a persistent feature near zero Doppler shift, 
which is coronal rain in the foreground of the eruption.
Because it is captured by the slit near the apexes of coronal loops 
where such cooling condensation is initially formed \citep[e.g.,][]{Antolin.coronal-rain.2010ApJ...716..154A, 
LiuW.Berger.Low.flmt-condense.2012ApJ...745L..21L, FangX.XiaC.MHD-coronal-rain.2013ApJ...771L..29F}, 
its small Doppler velocity, almost symmetrically distributed around zero, is expected
and consistent with our selection of the reference wavelength described in \sect{subsect_calib}.}
%

We can estimate the 3D velocity vector using simultaneous POS and Doppler velocities.
For example, during 02:30--02:45~UT, the two velocities (\fig{overview.eps}(f), red crosses and green curve) 
are almost the same, indicating an angle of $\sim$$45 \degree$ behind the POS. This is an upper limit,
because the eruption component~B has the highest redshift.
A more accurate estimate can be obtained during the late phase when the velocity distribution is relatively simple.
At 03:06:15~UT (see \fig{map-mg.eps}(h)), for instance,
the fallback material exhibits blueshifts increasing almost linearly with decreasing heights.
This is consistent with a similar trend of the POS velocity (white dashed line) obtained
from the last parabolic trajectory in \fig{overview.eps}(e) covering this time.
Specifically, the blueshift increases from $-72 \kmps$ to $-98 \kmps$ by 36\% over 
the POS height range of $43 \Mm$ ($59\arcsec$) within the \iris\ FOV. 
This percentage change is similar to the 33\% increase of the POS velocity
from $-137 \kmps$ at 03:03:32~UT to $-182 \kmps$ at 03:08:04~UT over the same height range. 
These two pairs of velocities give a consistent angle of $\arctan(72/137) \approx \arctan(98/182) = 28 \degree$
between the velocity vector and POS. 
This provides additional support for our interpretation of the late-phase blueshift as evidence of material falling back,
and implies that the fallback trajectory within this height range is oriented at a constant angle from the POS.

\revise{This parabolic space--time trajectory yields a downward POS acceleration 
of $165 \mpss = 0.60 g_\Sun$, which translates to a true 3D acceleration of 
$0.60 g_\Sun / \cos(28 \degree) = 0.68 g_\Sun$, where $g_\Sun=274 \mpss$ is the solar surface gravitational constant.
Note that Cut~1 (along which this trajectory is measured) is only $10 \degree$ from the POS projection		
of the radial direction and the $28 \degree$ angle of the velocity vector from the POS is close to the
$31 \degree$ angle of the behind-the-limb location of the source flare. Therefore, this fallback path is likely
close to the local vertical at the source%
 \footnote{The exact 3D path of the fallback material is difficult to infer in this case,
 because of the inadequate cadence of \stereo-A 	
 and thus the unknown landing site behind the limb.
 }
and the effective gravitational acceleration along it is close to $g_\Sun$,
greater than the measured $0.68 g_\Sun$.		
Such a less-than-free-fall acceleration, indicating cancellation of gravity by some upward force, 
is comparable to those (mostly measured in POS) of fallback material in chromospheric jets
\citep[e.g.,][]{LiuW.CaJet1.2009ApJ...707L..37L}
and of coronal rain \citep[e.g.,][]{Schrijver.coronal-rain.2001SoPh..198..325S,
Antolin.coronal-rain-mass-cycle.2012ApJ...745..152A},
but somewhat higher than those of downflow threads in quiescent prominences
\citep[e.g.,][]{vanBallegooijen.Cramer.tangled-field-promin.2010ApJ...711..164V,
ChaeJ.promin.threads.descend-knots.2010ApJ...714..618C,
LiuW.Berger.Low.flmt-condense.2012ApJ...745L..21L, 
LowBC.condenseP2.curren-sheet.2012ApJ...757...21L}.}  

\begin{figure*}[thbp]      
 \epsscale{1.1}
 \plotone{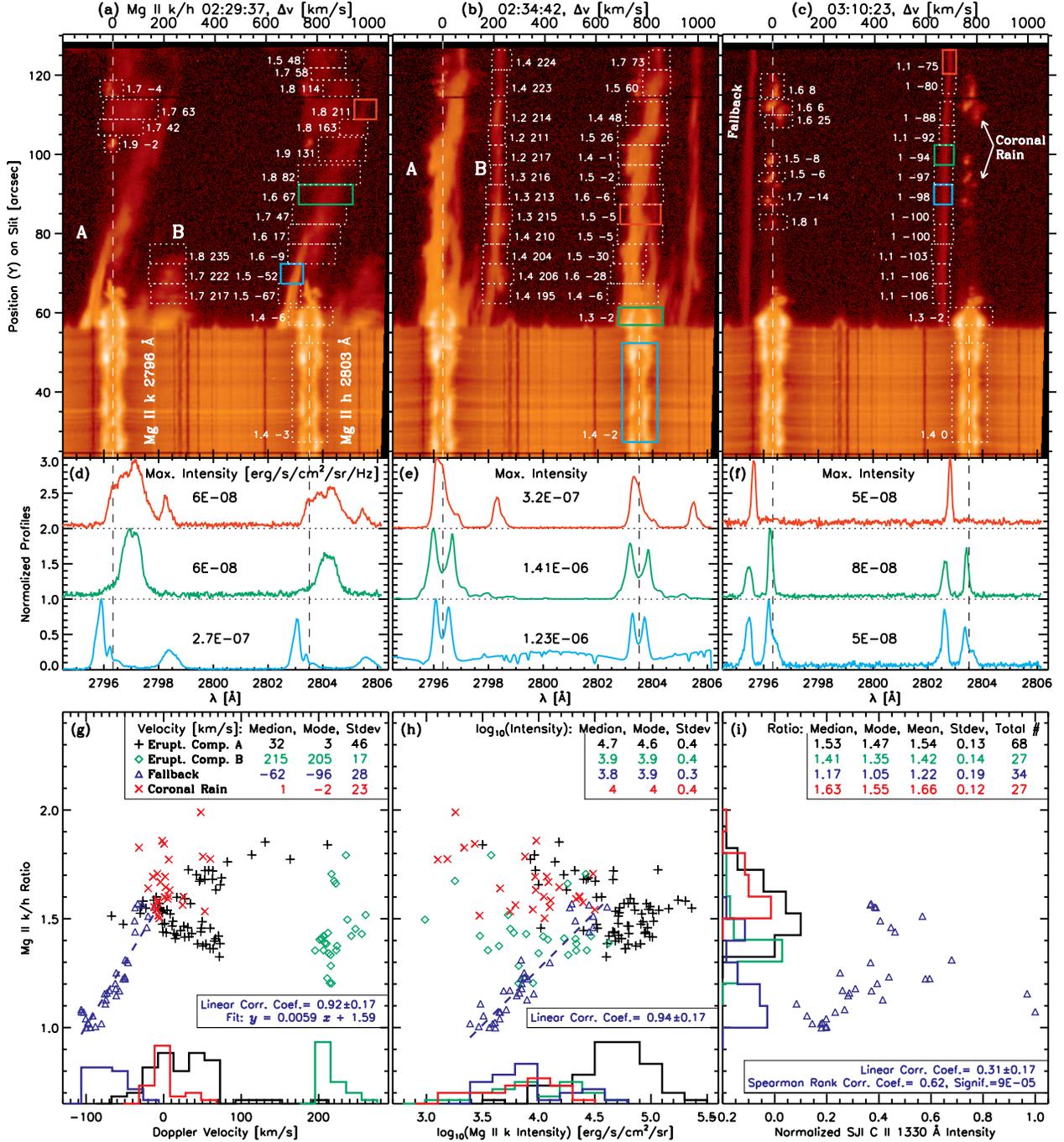}	
 \caption[]{\small
 Analysis of the \mgiik~2796~\AA\ to h~2803~\AA\ integrated line intensity ratio.	
   (a)--(c) \mgiik\ and h spectra at three selected times,	
 showing the primary (A) and secondary (B) eruption components in (a) and (b) and the fallback material together with
 the foreground coronal rain in (c). The boxes mark individual features selected for line ratio analysis,
 each labeled with the corresponding ratio and line-centroid Doppler velocity (in $\kmps$).
 The vertical dashed lines indicate the rest wavelengths.
   (d)--(f) Spectra averaged over the height ranges of selected boxes in corresponding colors shown above.
 Each profile is normalized and vertically shifted by multiples of unity.
 The number indicates the maximum of each profile in units of [$\erg \ps \pcms \psr \pHz$].
   (g) k/h intensity ratio vs.~line-centroid Doppler velocity from eight different times 
 (including those of (a)--(c)) during the course of the event,
 as marked by the plus signs in \fig{overview.eps}(h), at the same raster position. 
 The results are color-coded for four different features:
 the primary (A; black plus signs) and secondary (B; green diamonds) eruption components,
 fallback material (blue triangles), and coronal rain (red crosses).
 Their Doppler velocity histograms are plotted at the bottom, with the median, mode, and standard deviation
 values shown on the top. The fallback material exhibits a linear correlation between the two quantities,
 to which a linear fit is shown as the dashed line. 
   (h) Same as (g) but for the logarithmic k line intensity as the $x$-axis. 	
   (i) Histograms of the k/h intensity ratios of the four features plotted against the $y$-axis.
 The fallback material k/h ratio is also plotted vs.~the normalized
 SJI \cii~1330~\AA\ intensity on $x$-axis, showing a comparably weak, but noticeable correlation.
 } \label{ratio.eps}
 \end{figure*}

\revise{Another interesting feature of the fallback material is its narrow line width, which 
continues decreasing with time, as shown in \fig{overview.eps}(g). 
During the late phase after 03:10~UT (e.g., \figs{map-mg.eps}(h) and \ref{ratio.eps}(c)), its $1/e$ nonthermal line width
is only on the order of $10 \kmps$, nearly 50\% that of the foreground coronal rain at loop apexes.
This indicates that such material falls along streamline trajectories with very little
velocity scatter or nonthermal broadening. 
We speculate that this could be due to the lack of line broadening agents,
such as small-scale \Alfven\ waves or turbulence, which likely accompany the impulsive eruption earlier
but have diminished substantially ever since.}  

\subsection{Helical Structure and Torsional Motions}
\label{subsect_helix}


%
%
 \begin{figure*}[thbp]      
 \epsscale{1.1}
 \plotone{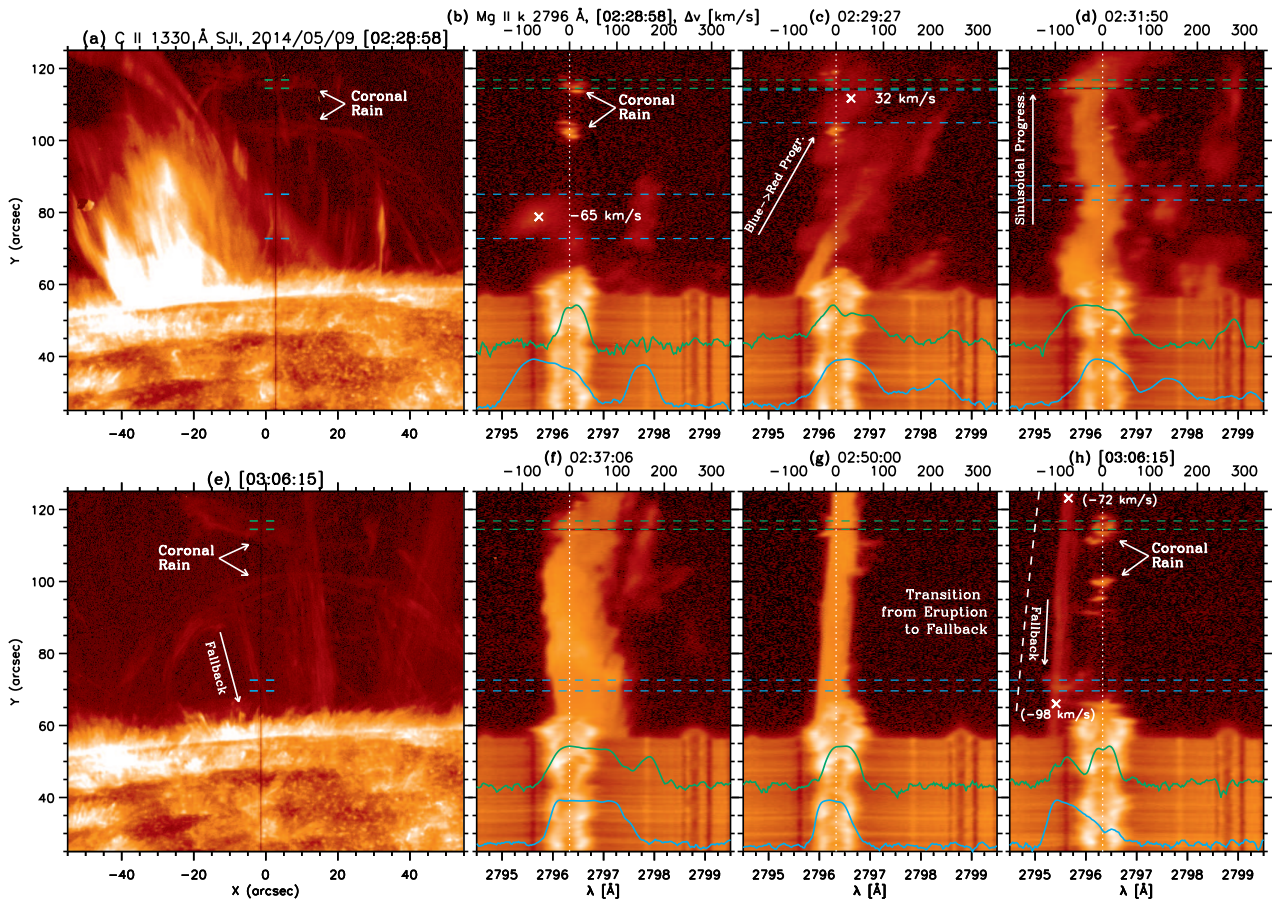}	
 \caption[]{	
 Same as \fig{map-3w.eps} but for the \mgiik\ 2796~\AA\ line alone at six selected times showing the detailed
 spectral evolution (see Animation~3). The corresponding times between the SJIs and
 spectra are shown in brackets.
  (b) and (c) show the blue- to redshift progression (e.g., from $-65$ to $32 \kmps$) with time and height of a single feature,
 while (d) shows an upward sinusoidal progression across the entire slit, both indicative
 of helical motions during the eruption.
  (f) and (g) show the two eruption components (A and B, see \fig{map-3w.eps}) gradually merging into one,
 which then decreases in line width with time and evolves from red- to blueshifts, while maintaining
 its generally constant slope with height.
   (h) Blueshifted spectrum of the falling material with narrow line widths during the late phase.
 The cross signs mark its line centroids at the top and bottom of the off-limb region.
 The white dashed line is the POS velocity as a function of distance ($y$-axis) 
 from the last parabolic fit shown in \fig{overview.eps}(e), which covers this time and the \iris\ FOV.
 Like the blueshifts, it exhibits a similar growth with decreasing heights,  
 indicating a true 3D downward acceleration of the falling material.
 } \label{map-mg.eps}
 \end{figure*}
%


Helical structures have been commonly seen in 	
imaging or spectroscopic observations of solar eruptions, including CMEs
\citep[e.g.,][]{KohlJ.UVCS.obs.high.corona.2006A&ARv..13...31K}, 
prominence eruptions \citep{KolevaK.AIA-1st-light.prom.erupt.helicity.2012A&A...540A.127K}, 
and jets or surges \citep{Canfield.surge-jet1996ApJ...464.1016C,
LiuW.CaJet2.2011ApJ}.		
We found compelling new evidence in this event, with the combination of 
high-resolution SJIs and spectra offering critical clues to the 3D structure. 

In the POS, AIA images show cork-screw shaped threads, especially early in the eruption
when the morphology is relatively simple (e.g., \figs{overview.eps}(a) and (b)).
Their temporal evolution is manifested in the
multiple sinusoidal tracks shown in \fig{overview.eps}(i), an \iris\ \cii~1330~\AA\ SJI space--time plot 
obtained from Cut~2, perpendicular to the eruption axis. In contrast, during the late phase, 
e.g., after 03:00~UT, the falling material shows no longer sinusoidal, but essentially flat tracks.
This suggests that the pre-stored magnetic helicity or twists have been transported into the heliosphere
by the eruption and thus the falling material returns as streamline flows along untwisted field lines,
similar to those in previously reported helical jets \citep[e.g.,][]{LiuW.CaJet1.2009ApJ...707L..37L}.

Doppler measurements by \iris\ provide several clues for the interpretation
of the above POS observations as torsional/rotational motions, rather than transverse oscillations.
The \iris\ slit is oriented at $13 \degree$ clockwise from the central axis of 
the eruption, allowing it to cover both sides of the axis, i.e., with its lower portion
sampling the left-hand side and the upper portion sampling the right-hand side.
We find that early in the eruption, the initial blueshifts of some individual features 
switch to redshifts {\it with time} as they travel upward along the slit (see \fig{map-mg.eps}(a)--(c) 
and Animation~2 for an example).
Moreover, in {\it single spectral snapshots}, the primary (A) spectral component of the eruption at times 
show predominant blueshifts at lower heights that smoothly transition to redshifts at larger heights 
(e.g., \fig{map-3w.eps}(b) and \fig{map-mg.eps}(c)).
The combination of these observations indicates a {\it counter-clockwise rotation}
of the erupting material when viewed top-down. 
Note that sometimes both components~A and B show no sign of blueshifts but only redshifts
that generally grow with height. This can be explained by the fact that the eruption
occurs behind the limb and large redshifts due to outward radial motions are expected
and can reduce or even reverse the blueshifts caused by rotations.

%

We can further infer the handedness of the helices by combining POS and Doppler observations.
Early in the event, SJIs show material threads progressing transversely from left to right.
To be consistent with the above inferred counter-clockwise rotations with dominant blueshifts on the left,
such threads must be located on the {\it front side} of the eruption,
although the \cii~1330~\AA\ SJI emission is likely optically thin above the \cii\ limb 
and thus we are seeing both the front and back sides. 
Noting that these threads are oriented from lower-right to upper-left, we conclude that 
they are {\it left-handed helices}. Their counter-clockwise rotations are thus consistent with 
relaxation or unwinding, rather than tightening, motions of the helices,
which are expected for such an eruption.

We also found sinusoidal spectral variations along the slit and their progression toward greater heights with time, 
best seen during 02:31:12--02:32:19~UT (\fig{map-mg.eps}(d)). This may be evidence of (multiple)
small-scale ($\sim$$10 \Mm$ across) helical sub-structures within the overall eruption,
which are also evident in SJIs (e.g., see the unfolding feature near 02:45~UT in Animation~2).
%
%

\subsection{\mgiik/h Line Ratio and Doppler Dimming}
\label{subsect_mg2}



\subsubsection{\mgiik/h Line Ratio}
\label{subsect_ratio}

The \mgiik~2796~\AA\ and h~2803~\AA\ lines and their intensity ratio can provide useful plasma diagnostics
\citep[e.g.,][]{LeenaartsJ.IRIS.Mg.paper1.2013ApJ...772...89L, PereiraTiago.IRIS.Mg.paper3.2013ApJ...778..143P}.
In general, this k/h ratio is expected to be $\sim$2 for optically thin, collisionally excited (thermal) emission.
Previously reported values include 2 for an active-region prominence \citep{VialJC.AR.prom.MgII.k/h.ratio=2.1979SoPh...61...39V},
1.7 for a quiescent prominence \citep{VialJC.quiescent.prom.MgII.k/h.ratio=1.7.1982ApJ...253..330V},
and 1.33--1.35 for quiescent prominences recently observed by \iris\ 
\citep{HeinzelP.IRIS.MgII.2014A&A...564A.132H, SchmiederB.IRIS.prom2014A&A...569A..85S}.

As shown in \fig{ratio.eps} (top), we identified various corresponding features 
in both the k and h lines, integrated their continuum-subtracted intensities, and obtained their ratios,
which are labeled together with the line-centroid Doppler velocities.
The mid row in \fig{ratio.eps} shows some sample spectra at selected positions.		
Unlike their on-disk or chromospheric counterparts, 
the off-limb \mgiik\ and h lines of both erupted material and coronal rain
have no central reversal, similar to those in \iris\ observed prominences 
\citep{HeinzelP.IRIS.MgII.2014A&A...564A.132H, SchmiederB.IRIS.prom2014A&A...569A..85S},
suggestive of a low pressure or thickness and optically-thin regime.
\revise{Because the oscillator strength of the k line is twice
that of the h line, the k line always has a larger integrated intensity.} 	
However, for the fallback material (\fig{ratio.eps}(f)), 
the k line has a comparably lower peak intensity.

We repeated this analysis for a total of 156 features selected from eight spectra 
through the course of this event. The resulting intensity ratio is plotted against
the line-centroid Doppler velocity and k line intensity (\fig{ratio.eps}(g) and (h)),
categorized for four types of features: 
the primary (A; black plus signs) and secondary (B; green diamonds) eruption components,
fallback material (blue triangles), and coronal rain (red crosses).
As a reference, the k/h ratios of the on-disk quiet Sun and chromosphere at the limb have 
medians and standard deviations of $1.38 \pm 0.02$ and $1.34 \pm 0.02$, respectively.
The coronal rain has the highest k/h ratio of $1.63 \pm 0.12 $, the lowest Doppler velocity nearly zero,
and a moderate intensity. Eruption component~A has a moderate k/h ratio of $1.53 \pm 0.13 $, 
a small Doppler velocity, and the highest intensity of $ 10^{4.7} \erg \ps \pcms \psr$.
Eruption component~B has a somewhat smaller ratio of $1.41 \pm 0.14 $, 
the highest Doppler velocity of $>$$200 \kmps$ redshifts, and a moderate intensity
about 6 times smaller than that of component~A.
The fallback material has the lowest k/h ratio of $1.17 \pm 0.19$, a moderate blueshift,
and the lowest intensity.

The k/h ratios of the fallback material are particularly interesting, yet puzzling.	
During the late phase of the event, as progressively less material is present off-limb and as the
intensity drops with time, one would expect the \mgiik\ and h emission to approach the optically-thin regime
and thus a k/h ratio close to 2. However, the $1.17$ median ratio is surprisingly small,
the lowest among all reported values for prominences and even less than those of the
quiet-Sun disk ($1.38 \pm 0.02$) and chromosphere ($1.34 \pm 0.02$). 
More interestingly, the k/h ratio $R_{\rm k/h}$ is highly correlated
with both the line-centroid Doppler velocity $v_{\rm D}$ and the k line intensity $I_{\rm k}$,
with linear correlation coefficients of $0.92 \pm 0.17$ and $0.94 \pm 0.17$, respectively.
A linear regression gives $R_{\rm k/h}= 0.0059 v_{\rm D}/(100 \kmps) + 1.59$, leading to a ratio of unity
at $v_{\rm D}=-100 \kmps$.

We also obtained \cii~1330~\AA\ SJI intensities averaged within the Y-direction ranges 
corresponding to those boxes defined in the spectra of the fallback material (e.g. \fig{ratio.eps}(c))
and within an X-direction range of $14 \arcsec$ covering all the 8-step raster positions.
They show a comparably weak but noticeable correlation with the k/h ratios. 	
This is perhaps not surprising because of the similar temporal trends
of the \cii~1330~\AA\ SJI and \mgiik\ line intensities, as shown in \fig{overview.eps}(d).

\subsubsection{Doppler Dimming}
\label{subsect_dim}

\referee{Compared with the eruption components~A and B with a complex morphology and dynamic evolution 
but lack of correlation in \figs{ratio.eps}(g) and (h), the fallback material has a relatively
simple morphology in a relaxed, post-eruption state and a well-defined correlation between
the \mgiik/h ratio and the Doppler velocity or line intensity. 
In this section, we thus focus on the fallback material to explore the underlying physics
of such characteristics as the surprisingly low values of the k/h ratio.} 	
We note that the off-limb k and h lines have contributions from both radiative excitation
by the solar surface radiation and local collisional (thermal) excitation. 
Their relative importance depends on the pressure (density and temperature) of the material
and is estimated as follows.

We assume that the \mgii\ line emitting plasma is optically thin, 
an assumption valid only for thin, low-pressure prominence threads 
\citep[see Table~2 of][]{HeinzelP.IRIS.MgII.2014A&A...564A.132H}.
Then each plasma blob receives the full incident radiation
and emits both scattered ($I_{\rm sc}$) and thermal ($I_{\rm th}$) radiation. 
Regardless of the frequency redistribution process, 
$I_{\rm sc}$ is the product of the geometric dilution factor 	
and the chromospheric intensity,		
for which we adopt the line-center intensity of $3\E{-7} \erg \ps \pcms \psr \pHz$
from \citet{HeinzelP.IRIS.MgII.2014A&A...564A.132H}.
$I_{\rm th}$ is the product of the photon destruction probability $\varepsilon$ 
and the Planck function $B_\nu$ at the plasma temperature $T$
\citep[see, e.g.,][]{MihalasDimitri.stellar.atomsph.2nd.ed.1978stat.book.....M,
VialJC.2D.rad.trans.prom.lines.1982ApJ...254..780V}. 
$\varepsilon$ is proportional to the electron density $n_{\rm e}$ 
and steadily increases with $T$ (so does $B_\nu$). 
We find that 

\begin{enumerate}

\item
for a typical prominence density of $n_{\rm e} \sim 10^{11} \pcmc$ 	
\citep[e.g., see Table~4 of][]{Labrosse.prominence-review.2010SSRv..151..243L}
and a low temperature of $T=6\E{3} \K$, the radiation term $I_{\rm sc}$ dominates over the collisional term $I_{\rm th}$ 
by orders of magnitude, while 

\item
an increase in density above $10^{12} \pcmc$ 
and/or in temperature to $1.5 \E{4} \K$ would result in the collisional-term dominance.	

\end{enumerate}


%
 \begin{figure}[thbp]      
 \epsscale{1.15}	
 \plotone{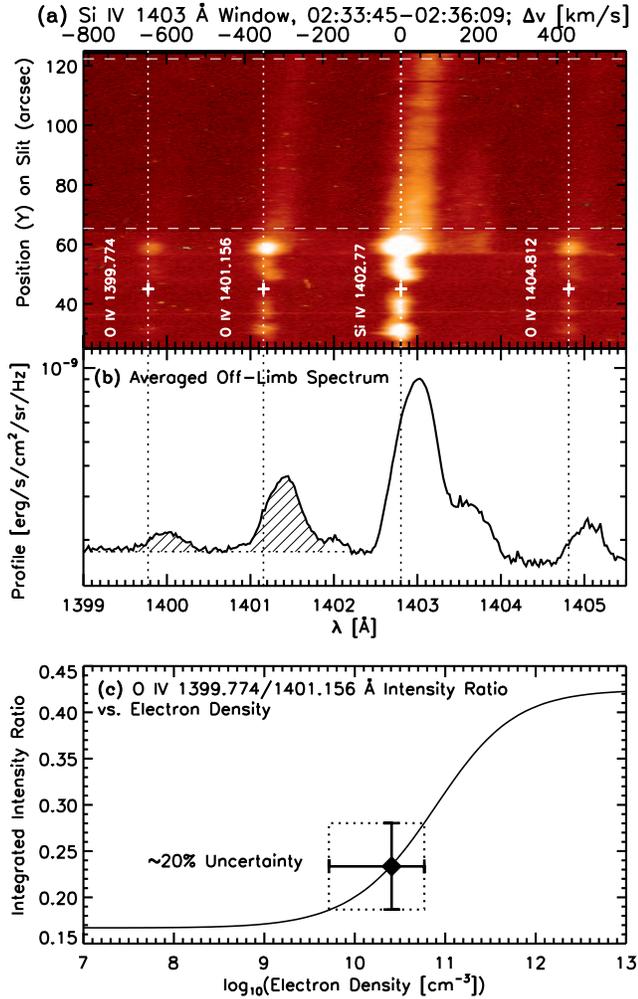}	
 \caption[]{	
 \referee{Density estimate using the \oiv~1399.774/1401.156~\AA\ line ratio.}
  (a) Spectrum of the \ion{Si}{4} 1403~\AA\ window averaged over 16 spectra of two raster scans
 during 02:33:45--02:36:09~UT.
  (b) Spectrum averaged over the off-limb region between the two horizontal dashed lines in (a).
 The \oiv\ 1399.774 and 1401.156~\AA\ line intensities of the primary eruption component (A)
 integrated over the two hatched areas gives a ratio of 0.234,
 shown as the diamond in (c). 	
 The curve in (c) is the CHIANTI-predicted line ratio as a function of electron density.
 } \label{dens.eps}
 \end{figure}
%
%
Determining the relative importance of $I_{\rm sc}$ and $I_{\rm th}$	
for the specific event under study requires the knowledge of the plasma density.	
To this end, we utilized the density sensitive \oiv~1399.774/1401.156~\AA\ line ratio.
In order to increase the signal-to-noise ratio, we first obtained a spectrum (see \fig{dens.eps}(a)) by averaging
two raster scans when these lines are relatively bright and overlaps due to 
Doppler shifts are relatively small. We then averaged all off-limb spectra to obtain
the profile shown in \fig{dens.eps}(b), which then gives a continuum-subtracted, 
integrated \oiv~1399.774/1401.156~\AA\ intensity ratio of 0.234
for the primary (A) eruption component.
According to the CHIANTI~v7.1 database \citep{LandiE.CHIANTI.XIII.v7.1.2013ApJ...763...86L},
this ratio translates to a predicted density of $n_{\rm e}=2.6 \E{10} \pcmc$ (see \fig{dens.eps}(c)).
Its uncertainty originates mainly from line blending, which is estimated as follows:
(i) We found that the \oiv~1399.774~\AA\ line is blended by the \feii~1399.962~\AA\ line 
at an upper-limit 6\% level. 
\referee{To arrive at this order-of-magnitude estimate, we assumed that the \feii\ and \mgiik\ lines
are emitted by the same optically-thin plasma of a uniform differential emission measure (not necessarily true). 
We used the CHIANTI routine \texttt{ch\_synthetic.pro} and assumed a constant pressure
(corresponding to $n_{\rm e}= 10^{11} \pcmc$ and $T=10^4 \K$ for typical prominence conditions)
to synthesize their respective contribution functions $G(T)$, 
which were then multiplied by the chromospheric abundance
and the \iris\ response function (from \texttt{iris\_get\_response.pro})
and integrated over a temperature range of $10^4$--$10^5 \K$ covering the $G(T)$ peak
to yield their line intensities, $I_{\rm Fe \, II}$ and $I_{\rm Mg \, II \, k}$.		
With the predicted ratio of $I_{\rm Fe \, II} / I_{\rm Mg \, II \, k} = 3.7\E{-4}$
and the observed \mgiik\ intensity, we estimated the \feii~1399.962~\AA\ intensity,
which turned out to be $\leq$6\% of the observed \oiv~1399.774~\AA\ intensity.}  
(ii) We could not identify any \si~1401.514~\AA\ line emission beyond the quiet-Sun disk and thus its blend to the 
\oiv~1401.156~\AA\ line is considered negligible. 
Nevertheless, we assumed a conservative 20\% uncertainty for the line ratio, which gives a density range
of $5.2\E{9}$\,--\,$5.9\E{10} \pcmc$. 
\revise{This density pertains to the warm plasma at the \oiv\ formation temperature of $\sim$$10^{5.2} \K$,
which, in this case, is most likely located in the so-called 
prominence--corona transition region \citep[PCTR;][]{ParentiS.Via.PCTR.2014IAUS..300...69P}.
In the cold prominence core 	
at the \mgiik\ and h formation temperature of $\sim$$10^{4} \K$,
the density could be $10$ times higher (assuming pressure equilibrium), 
but still smaller than the extreme density of $\geq$$10^{12} \pcmc$ 
required for the collisional-term dominance in the low temperature regime as mentioned above.
Therefore, we conclude that the \mgiik~and h emission 
of the erupted material is dominated by radiative excitation.}  


%
 \begin{figure}[thbp]      
 \epsscale{0.95}	
 \plotone{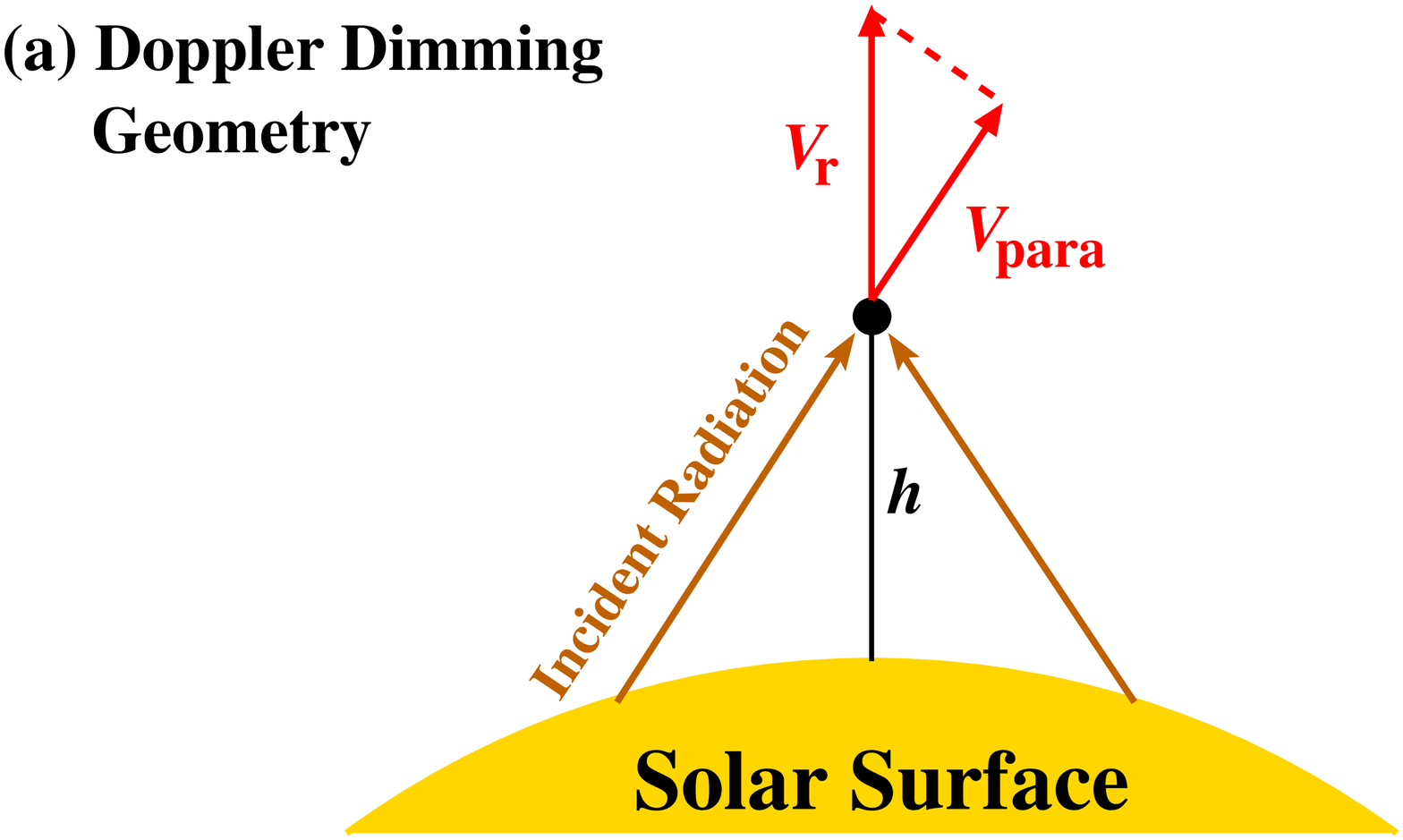}
 \epsscale{1.1}	
 \plotone{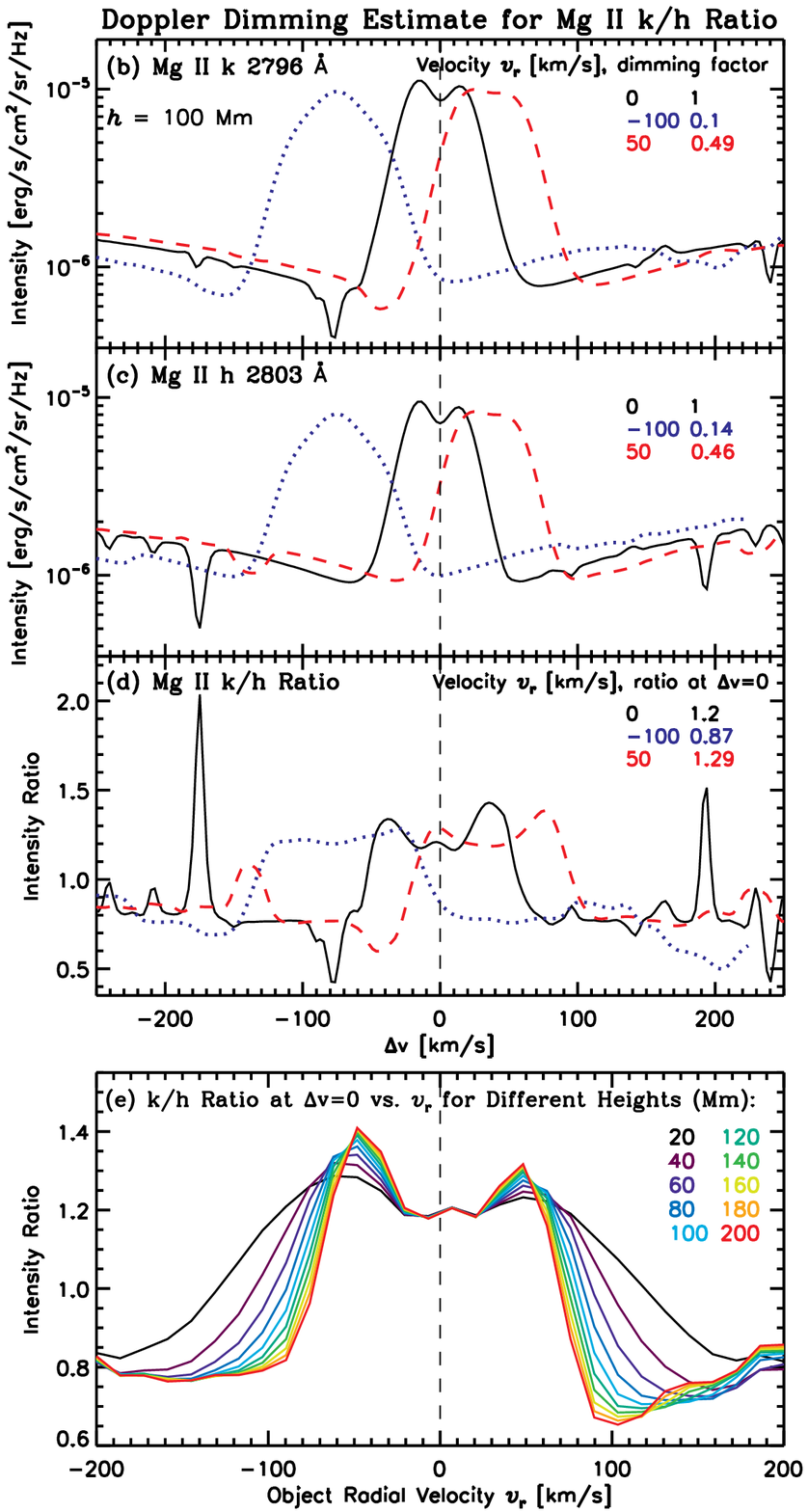}	
 \caption[]{	
 \referee{Empirical modeling of the \ion{Mg}{2}~k/h line ratio due to Doppler dimming.}
  (a) Schematic geometry for Doppler dimming estimate shown on the right.
  (b) and (c) \ion{Mg}{2}~k 2796~\AA\ and h 2803~\AA\ line profiles (centered at their respective
 rest wavelengths) averaged over the entire visible solar disk as seen by a radially moving object 
 at three velocities ($v_{\rm r}= -100$, 0, $50 \kmps$) and a fixed height $h=100 \Mm$,
 with their corresponding k/h ratios shown in (d).
  (e) \ion{Mg}{2}~k/h line ratio at the line center ($\Delta v= 0$), as shown in (d), as a function of the object's
 radial velocity $v_{\rm r}$ for a range of color-coded heights (20--200~Mm). A positive correlation is present
 for falling material ($v_{\rm r}$$<$0), in qualitative agreement with the correlation shown in \fig{ratio.eps}(g).
 } \label{model.eps}
 \end{figure}
%
For radiatively excited emission from moving objects,
there is a {\it Doppler dimming} effect in which the incident line emission from the solar surface 		
is Doppler-shifted out of resonance. This effect has been extensively investigated for hydrogen and helium lines
\citep[e.g.,][]{HyderC.LitesB.Ha.LymA.Doppler.dim.1970SoPh...14..147H,
HeinzelP.prom.H-line.Doppler.dim.1987SoPh..110..171H,
GontikakisC.prom.H.lines.Doppler.dim.1997A&A...325..803G,
LabrosseN.prom.He.line.Doppler.dim.2007A&A...463.1171L, 
LabrosseN.prom.AIA.He.304A.Doppler.dim.2012A&A...537A.100L}
and recently modeled for the \mgiik\ and h lines \citep{HeinzelP.IRIS.MgII.2014A&A...564A.132H}.
Note that the wings of the chromospheric k and h lines are somewhat different. Therefore, when a moving object
resonantly absorbs the Doppler-shifted wings followed by de-excitation radiation, 
the emergent k/h ratio can depend on the Doppler velocity.
This may potentially explain their observed correlation.		

As a proof-of-concept exercise, we model the \mgiik/h ratio due to Doppler dimming as follows. 
Again, we assume that the erupted material is optically thin and ignore collisional excitation
according to the above analysis. Thus the \mgiik\ and h emission 
is simply resonant scattering of the incident chromospheric radiation, with no radiative transfer taking place.
We also assume that the solar surface is a sphere producing uniform radiation, for which we adopt
an \iris-observed quiet-Sun \mgiik\ and h spectrum, and ignore limb darkening or brightening. 
For an object at a height $h$ moving at a velocity $v_{\rm r}$ in the radial direction, we calculate its incident radiation
by integrating the Doppler-shifted radiation from the entire visible solar surface, 
as schematically shown in \fig{model.eps}(a). This spatial average can alter the shape of the
absorbed line profile, because the radial velocity $v_{\rm r}$ has different
parallel components $v_{\rm para}$ projected along incident rays from different directions 
that determine the Doppler shifts. As an example, 
\fig{model.eps}(b) and (c) show the average \mgiik\ and h profiles incident on 
an object moving at different velocities at a height of 100~Mm,  
which is near the lower limit of the height range of the \iris-observed prominence material 
according to its behind-the-limb source location. 
At $\gtrsim$$50 \kmps$, the central reversals of both lines disappear.
Because Mg is a heavy element, 	
the thermal broadening is expected to be small
and we assume a delta-function profile for the scattering agent (moving object).
Thus the scattered emission is essentially the incident radiation at the line center ($\Delta v=0$),
as marked by the vertical dashed line. The ratio of the k and h intensity at this 
position, as shown in \fig{model.eps}(d), would be equivalent to the observed line ratio.
For a stationary object ($v_{\rm r}=0$), this k/h ratio is $R_{\rm k/h}=1.20$.
For a moving object, the two lines are dimmed by different factors.
At $v_{\rm r}= -100 \kmps$, for example, the k line center intensity
is reduced by a factor of 0.10, while the h line by 0.14,%
leading to $R_{\rm k/h}=1.20 \times (0.10/0.14) = 0.87$.
Such dimming factors are on the same order of magnitude
as those modeled for a moving prominence \citep{HeinzelP.IRIS.MgII.2014A&A...564A.132H}.

Repeating this for different height $h$ and velocity $v_{\rm r}$ values,
we obtained the dependence of the k/h ratio $R_{\rm k/h}$ on these two quantities,
as shown in \fig{model.eps}(e).		
These curves have a butterfly shape and generally decrease with increasing $v_{\rm r}$ or $h$. 
There is a positive correlation between $R_{\rm k/h}$ and $v_{\rm r}$ 
near $v_{\rm r}= -100 \kmps$ ($<$0 for falling material), which qualitatively agrees with the 
observed correlation shown in \fig{ratio.eps}(g).
A blue-red asymmetry is present, with smaller k/h ratios for radially outward velocities ($v_{\rm r}$$>$0),
because of the asymmetry in the \mgiik\ and h wings.
For lower heights, the butterfly curves have a broader top portion 
at values close to that at $v_{\rm r}=0$,
because relatively more incident radiation originates from large angles
away from the local radial direction at the object. Such radiation is less Doppler-shifted
due to the smaller $v_{\rm para}$ component of the assumed radial velocity projected along those rays.


Note that the eruption component~B is about $200 \kmps$ higher in Doppler velocity
and a factor of 6 fainter in intensity than component~A, and even fainter than the coronal rain.		
We suggest that Doppler dimming may contribute to its faintness, among other possible factors, 
such as a smaller emission measure. 
\revise{We also note that the gradual fading of the \cii~1330~\AA\ SJI intensity 
during the late phase (see \fig{overview.eps}(d) and online Animation~2) coincides with the increasing blueshifts
and POS downflow velocities. We speculate that this fading may be due, at least in part, to
Doppler dimming of the \cii\ lines in a similar manner.}  
\referee{In contrast, Doppler dimming has very little effect on the coronal rain detected
near the loop apexes because of its small velocities.}

\section{Concluding Remarks}    
\label{sect_conclude}


\subsection{Summary}
\label{subsect_summary}

We have presented the first \iris\ observations of a fast prominence eruption associated
with a CME and an equivalent X1.6 flare on the far side of the Sun.
We summarize our major findings as follows.

\begin{enumerate}	

\item	

The ejected material is detected in bright lines at chromospheric to transition-region temperatures
($10^{4}$--$10^{5.2} \K$). We find no obvious delay among lines at different temperatures,
suggesting {\it no detectable temperature changes} (see \sect{subsect_tem}).

\item	

The prominence eruption has a maximum POS velocity of $\sim$$1200 \kmps$ and redshift of $460 \kmps$, while
the white-light CME has a maximum POS speed of $1300 \kmps$ (\sect{subsect_vel}).
These values are near the {\it higher end} of the velocity distribution reported for prominence eruptions. 
\revise{In contrast, the Doppler velocities of prominence material in the slower CME interior
detected by \soho/UVCS are on average only $\sim$10\% of the POS speeds of 
the CME leading fronts seen by \soho/LASCO \citep{GiordanoS.UVCS.CME.catalog.2013JGRA..118..967G}.}  

\item	

The erupting material exhibits a cascade with time toward lower velocities,
with a transition from upward ejection to downward fallback consistently manifested
in {\it both POS velocities and Doppler shifts} (red to blue drift). 
During the fallback phase, at a given time, larger blueshifts are found at lower heights,
consistent with the downward acceleration of the falling material;
at a given height, the blueshift increases with time, consistent with the fact that
material falling later returns from greater heights and is originally ejected at larger initial velocities earlier.
\revise{A typical fallback trajectory has a less-than-free-fall, true 3D acceleration of $0.68 g_\Sun$,
comparable to the values reported for falling ejecta and coronal rain.
The fallback material exhibits a progressively narrower line width down to $\sim$$10 \kmps$,
indicative of streamline flows (see \sect{subsect_vel}).
}  

\item	

There are {\it two components} of erupted material in Doppler velocity: a primary (A), bright component
of comparably small Doppler shifts and a secondary (B), faint, highly redshifted component.
The two components are separated by $\sim$$200 \kmps$, suggestive of a hollow, 
rather than solid, cone shape, in which the material is distributed (\sect{subsect_vel}).
We also suggest that stronger Doppler dimming of component~B can contribute to its relatively smaller intensity.

\item	

The combination of a blue- to redshift transition with height along the slit
and cork-screw shaped threads and sinusoidal transverse motions imaged in the POS
indicates that the eruption involves a {\it left-handed helical structure} 
undergoing counter-clockwise (when viewed top-down) unwinding/relaxation motions
(\sect{subsect_helix}).
Such a handedness originating from the southern hemisphere (AR~12051) is opposite to 
the dominant hemispheric rule of magnetic helicity for active regions \citep[e.g.,][]
{PevtsovA.helicity.hemisph.rule.1995ApJ...440L.109P, WangY-M.helicity.hemisph.rule.strength.2013ApJ...775L..46W}.


\item	

We find a wide range of \mgiik/h line intensity ratios.
The foreground coronal rain has the highest median ratio of $1.63$, 
the eruption components~A and B have intermediate values of $1.53$ and $1.41$, respectively,
while the fallback material has the lowest value of $1.17$, the {\it smallest ever reported}.
In addition, the k/h ratio of the fallback material is {\it strongly correlated} with 
the Doppler velocity and line intensity, with linear correlation coefficients 
at $>$$5\sigma$ levels (\sect{subsect_ratio}).

\item	

The low density $n_{\rm e}< 10^{12} \pcmc$ of the prominence core
inferred from the \oiv~1399.774/1401.156~\AA\ line ratio implies 
that the observed \mgiik\ and h emission is dominated by radiative (rather than collisional) excitation.
Doppler dimming is thus expected to alter the emergent emission profile
from moving objects.	
Our simple back-of-envelope calculation demonstrates that this effect may qualitatively explain the observed
\mgiik/h ratios, especially the surprisingly low values found in the fallback material and their correlation with Doppler velocities
(\sect{subsect_dim}).

\end{enumerate}		

\subsection{Discussion}
\label{subsect_discuss}


New \iris\ observations, such as these presented here, have opened a new
diagnostic window to study CMEs/prominence eruptions. 	
\revise{For example, once the thermodynamic parameters establish that the
\mgiik\ and h emission is dominated by resonance scattering, 
the k/h ratio can provide, through the Doppler dimming factor, 
critical information about the radial velocity, height, and 3D geometry of the ejected material.
Similar techniques have already been used for diagnosing solar wind and CME speeds
in the outer corona \citep[e.g.,][]{KohlJ.solar.wind.v.by.Doppler.dimming.1982ApJ...256..263K,
RaymondJ.UVCS.CME.vel.by.Doppler.dimming.2004ApJ...606L.159R}.}  

Another novel feature revealed by \iris\ is that
the erupting and returning material behind the limb both appear
at spicule heights within the chromosphere in all bright lines at large Doppler shifts
(see \fig{map-3w.eps}), alongside their corresponding rest-wavelength chromospheric emission at the foreground limb.
These lines include the \ion{Mg}{2}~k and h and \ion{C}{2} lines,
which have central reversals 	
in the chromospheric emission because of high opacity, 
but have no reversal in the Doppler-shifted, off-limb components indicating a largely optically-thin regime.
In other words, \iris\ {\it ``sees through" the chromosphere \it at the limb} in certain Doppler-shifted lines 
of otherwise optically-thick chromospheric emission. 
This is equivalent to detecting emission at different wavelengths 
from different LOS depths of the on-disk atmosphere.
This capability of simultaneously detecting behind-the-chromospheric-limb, Doppler-shifted material
(which is usually not accessible to telescopes from the Earth view)
and the foreground chromospheric material of the same temperature 
provides a unique, novel diagnostic potential yet to be explored,
e.g., for partially occulted flares 
\citep[e.g.,][]{LiuW_2LT.2008ApJ...676..704L, KruckerS.occult-stat2008ApJ...673.1181K}.

The proof-of-concept estimate of Doppler dimming presented in \sect{subsect_dim} has its limitations
with assumptions that may not be valid in the real situation. It explains some, but not all observed
features including some high \mgiik/h ratio ($>$1.6) of the highly redshifted eruption component~B (\fig{ratio.eps}(g)).
However, its promise demonstrated in the qualitative agreement with the observations		
warrants more detailed NLTE modeling
\referee{with the inclusion of, e.g., a 2D or 3D geometry, velocity vectors for
Doppler dimming effects, and, as in \citet{HeinzelP.IRIS.MgII.2014A&A...564A.132H},
the presence of a prominence--corona transition region,
which we plan to pursue in the future.}
\revise{We also plan to use multi-instrument observations to infer the detailed
3D geometry and velocity vector as well as the escaping and returning mass fluxes.} 



\acknowledgments
{	
This work is supported by NASA contract NNG09FA40C (IRIS) and the Lockheed Martin Independent
Research Program. 	
W.L.~thanks Paola Testa and Lucia Kleint for help with \iris\ data analysis,		
 \iris\ planner Nicole Schanche	
for capturing this eruption, and Hui Tian and Adrian Daw for useful discussions. 
}






{\scriptsize

}

\end{document}